# Computer simulation of the Poisson's ratio of soft polydisperse discs at zero temperature

Jakub Narojczyk,  Krzysztof W. Wojciechowski

Institute of Molecular Physics, Polish Academy of Sciences
ul. M. Smoluchowskiego 17, PL-60179 Poznań, Poland
E-mail: kww@man.poznan.pl

Abstract:
A simple algorithm is proposed for studies of structural and elastic properties in the presence of structural disorder at zero temperature. The algorithm is used to determine the properties of the polydisperse soft disc system. It is shown that the Poisson's ratio of the system essentially depends on the size polydispersity parameter - larger polydispersity implies larger Poisson's ratio. In the presence of any size polidispersity the Poisson's ratio increases also when the interactions between the particles tend to the hard potential.

## 1. Introduction

It is expected that future 'intelligent materials' will combine various unusual properties, e.g. unusual electromagnetic properties with unusual elastic properties. Although the latter subject (elasticity) has been the field of human's investigation since the ancient times, many problems interesting both from the point of view of basic research and from the point of view of possible applications remain unsolved. One of such problems is the influence of various forms of disorder on the elastic properties of matter.

In the present note we concentrate on studies of one of the forms of disorder - the size polydispersity. The question we pose is: How the elastic properties of a system are modified when instead of being consisted of identical particles it is formed of particles having some distribution of sizes? Polydisperse systems have been recently intensively studied because of their role in various fields of science and technology[1,2]. In contrast to the phase diagram and structure, however, their elastic properties remain an open field.

The subject of our study is a two-dimensional polydisperse model system at zero temperature. The simulations done here were meant to provide additional data to the work described in [3] in which the investigations where carried out at positive temperatures. We concentrate on studies of the Poisson's ratio[4] which directly describes the deformation of materials under loading/unloading stress. The latter subject is related to increasing interest in, so called, auxetic materials, i.e. systems showing negative Poisson's ratio [5-9]. Studies of influence of various mechanisms on the Poisson's ratio can help in searching for or manufacturing real materials of this unusual property.

The paper in organized as follows. In section 2 the studied model is described. In section 3 some theoretical background is reminded. In section 4 the chosen method of simulations is sketched. In section 5 the results are presented. Conclusions are drawn in section 6.

## 2. The system studied

The system under study, shown in Figure 1 and further referred to as the *polydisperse soft disc system*, consists of soft particles that interact only with their nearest neighbours through the interaction potential of the form

$$u_{ij}(r_{ij}) = \left(\frac{d_1+d_2}{2r_{ij}}\right)^n, \qquad (1)$$

where $d_1$ and $d_2$ are the diameters of the interacting particles. The values of the diameters were generated [10] according to the Gauss distribution function with the fixed size polydispersity parameter, $\delta$, defined as

$$\delta = \frac{\left(\langle\sigma^2\rangle - \langle\sigma\rangle^2\right)^{1/2}}{\langle\sigma\rangle}. \qquad (2)$$

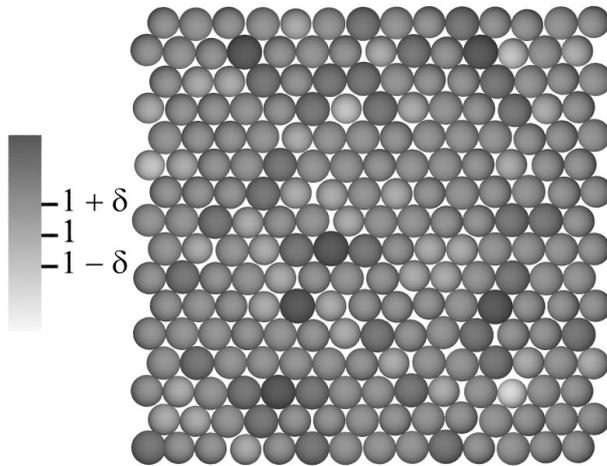

**Fig. 1**. Typical configuration of the studied system for N=224 particles; $\delta$ is the polydispersity parameter. Discs of different sizes are represented by different intensities of greyness.

It can be seen in Figure 1 that the soft discs form nearly hexagonal lattice for which the elastic properties are isotropic for small deformations [4]. It is worth to stress that when the exponent *n* tends to infinity, the above system tends to the static (i.e. zero temperature), polydisperse, hard disc system, studied in [3] at positive temperatures.

Our aim was to determine the dependence of the elastic constants and the Poisson's ratio of a given structure on the amount of disorder introduced into the system; the disorder was quantified by the parameter $\delta$. The data which will be shown below were obtained by computer simulation. Although the polydisperse soft disc system is very simple, it includes two main features of various real systems: repulsive forces at high densities and certain distribution of sizes of the interacting bodies. Hence, one can expect that it will supply some (qualitative, at least) information on the behaviour of more complex, real systems.

## 3. Theoretical analysis

As the studied system is isotropic from the point of view of its elastic properties, it can be described by only two elastic constants [4]: the bulk modulus, *B*, and the shear modulus, $\mu$. The elastic constants *B* and $\mu$ characterize the material's resistance against the changes of the volume and the shape, respectively. Both mentioned values must be positive in the range of the mechanical stability of the system.

The free energy of the deformed two-dimensional (2D) crystal exhibiting a 6-fold symmetry axis can be written as a function of the strain tensor components and the elastic constants in the form [4]

$$F = -p(\varepsilon_{xx} + \varepsilon_{yy}) + 2\lambda_{\xi\eta\xi\eta}(\varepsilon_{xx} + \varepsilon_{yy})^2 + \lambda_{\xi\xi\eta\eta}[(\varepsilon_{xx} + \varepsilon_{yy})^2 + 4\varepsilon_{xy}^2]. \qquad (3)$$

The values of the mentioned above elastic constants and pressure can be obtained by differentiation of the free energy with respect to the strain tensor components, see e.g. [11]

$$\begin{aligned}
\left.\frac{\partial F}{\partial \varepsilon_{xx}}\right|_{\varepsilon=0} &= p \quad, \\
\left.\frac{\partial^2 F}{\partial \varepsilon_{xx}^2}\right|_{\varepsilon=0} &= 4\lambda_{\xi\eta\xi\eta} + 2\lambda_{\xi\xi\eta\eta} \quad, \\
\left.\frac{\partial^2 F}{\partial \varepsilon_{xx}\partial \varepsilon_{yy}}\right|_{\varepsilon=o} &= 4\lambda_{\xi\eta\xi\eta} - 2\lambda_{\xi\xi\eta\eta} \quad.
\end{aligned} \qquad (4)$$

The bulk modulus, *B*, and the shear modulus, $\mu$, can be easily related [12] to the quantities used in the equation (3)

$$B = 4\lambda_{\xi\eta\xi\eta} \;,$$
$$\mu = 2\lambda_{\xi\xi\eta\eta} - p \;. \tag{5}$$

The Poisson's ratio for an isotropic 2D system can be expressed by the above elastic moduli as follows [12]

$$\nu = \frac{B - \mu}{B + \mu}. \tag{6}$$

It is worth to notice, that in the case of the discs of equal diameters ($\delta = 0$) one can obtain analytical formulae for the elastic constants

$$p_0 = n\sqrt{3}a^{-n-2} \;,$$
$$B_0 = \left(\frac{n}{2} + 1\right)p_0 \;,$$
$$\mu_0 = \left(\frac{n}{4} - \frac{1}{2}\right)p_0 \;, \tag{7}$$
$$\nu_0 = \frac{n+6}{3n+2} \;.$$

where $n$ is the power of the potential and $a$ is the distance between the centres of the particles.

### 4. The simulation method

The elastic properties with respect to the degree of polydispersity [3] were studied. For each of more than 1000 different structures the reference state of minimal energy was found. Then the deformation and the "measurement" took place.

Simulations where carried out based on the algorithm which searched for a configuration of minimum energy for a given shape of the system. The studied model was enclosed in periodic boundary conditions. Its total energy was a sum of the interactions of the form (1) between the nearest-neighboring particles (i.e. those whose Dirichlet polygons have a common side). In search for the minimum energy states, the program conducted a series of translational moves of particles; only those decreasing the energy of the system were accepted. Particles were being moved in direction of the total force acting on them, by an arbitrary vector $dr_i$. After acceptance/rejection of each move, the mentioned vector was being modified with respect to the change of the total system's energy: if the move led to increase/decrease of the energy the $dr_i$ was decreased/increased, respectively. The particles were being moved one-by-one in the order they were placed in the lattice. (Enumeration of the particles was arbitrary and was checked to have no influence on the application of the algorithm and its results.) To avoid trapping the system into a local minimum some "shaking" (random changes of the particle positions) was introduced. Each of the particles was being randomly displaced by a vector $dr_{rand}$ of random length and orientation, generated within a defined range of values. The program moved all the particles of the system as long as the system's energy was no longer changing within the given accuracy.

After calculating the energy of the system at equilibrium (equal to its free energy at zero temperature) the shape of the system was deformed by given strains for which the minimum energy configurations of particles were searched for. Then a numerical differentiation of the energy with respect to the deformations was used to determine the values of elastic constants $B$ and $\mu$ from which the Poisson's ratio of the studied system was obtained. The precise description of the algorithm is given in [13].

## 5. Results

The figures presented below summarize the simulation results. It can be seen that for large values of the exponent, $n$, the elastic properties of the system strongly depend on the polydispersity of the system. In particular, it can be seen in Figure 2 that when $n \to \infty$ both $B$ and $\mu$ grow rapidly with respect to the values obtained for systems composed of identical particles.

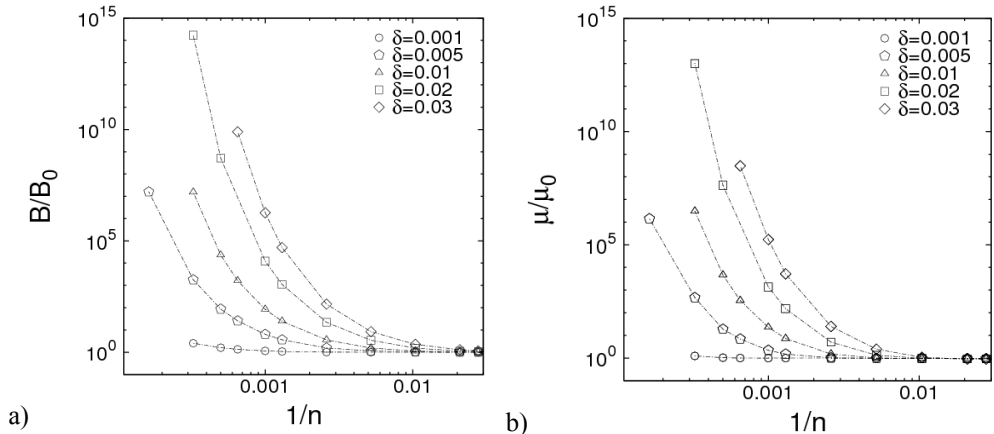

**Fig. 2**. The dependence of **(a)** the bulk modulus, $B$, and **(b)** the shear modulus, $\mu$, on the inverse of the exponent $n$. The doted lines are presented to guide the eye.

Analyzing the Figure 3 one can notice that by increasing the polydispersity of the system, its Poisson's ratio grows with respect to the periodic system of discs with identical diameters. The Figure suggests that when $n \to \infty$ then for any non-zero polydispertity the Poisson's ratio tends to +1, what is its maximum value possible for a two-dimensional isotropic system.

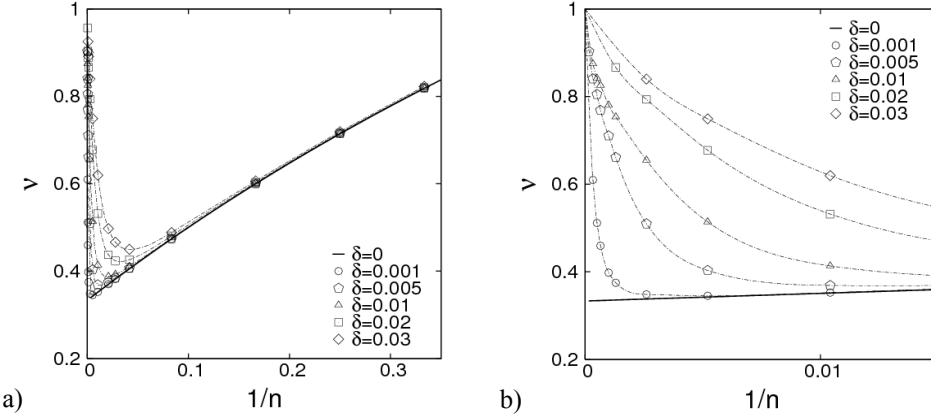

**Fig. 3.** The value of the Poisson's ratio against the inverse of the exponent *n* for two-dimensional soft discs with different values of the polydipersity parameter: (a) $n \geq 3$, (b) $n \geq 96$. The thick continuous line shows the exact result for $\delta=0$, see equation (7). The doted lines are presented to guide the eye.

Closing this section, we should add that the elastic properties of the polydisperse soft disc system at zero temperature are in very good agreement with Monte Carlo simulations of the polydisperse soft discs at *positive* temperatures, when the temperature tends to zero[14].

## 6. Conclusions

A simple and efficient algorithm to study static structures of polydisperse soft particles has been proposed. It was shown that the static disorder studied (the lack of periodicity of the particle positions caused by unequal sizes of the particles) has an essential influence on the value of the Poisson's ratio. Typically, for positive values of the polydispersity parameter, the bulk modulus, *B*, the shear modulus, $\mu$, and the Poisson's ratio, *v*, increase with increasing *n* when potentials become strongly repulsive, i.e. when the exponent *n* is large, what corresponds to the hard particle limit. These quantities increase also with increasing polydispersity parameter, $\delta$, in the same limit of large values of *n*.

The results obtained in this work show an excellent agreement with recent results obtained by Monte Carlo simulations of the soft polydisperse discs in the low temperature limit [3], which in turn agree very well with the results obtained for hard polydisperse discs [14] in the limit of high temperatures and large *n*. This confirms the hypothesis that the elastic properties of hard-body systems can be obtained by studying soft-body systems and taking the $n \to \infty$ limit. We plan to apply the present simulation method to three-dimensional systems of isotropic and anisotropic particles which were recently studied at positive temperatures. We expect that studies of very simple models, as the one described in the present paper, will help in understanding and predicting various behaviours observed in real systems. It is interesting to check to what extent such simple models approximate the properties of some real systems, like those described in [15].

## Acknowledgments

This work was supported by the (Polish) Committee for Scientific Research (grant No. 4T11F01023) and by the Centre of Excellence for Magnetic and Molecular Materials for Future Electronics within the European Commission contract No. G5MA-CT-2002-04049. The simulations were performed at the Poznań Supercomputing and Networking Center.